# Compact Two-dimensional RAID Level 6 Arrays Tolerating Quadruple Node Failures

Jehan-François Pâris
*Department of Computer Science*
*University of Houston*
*Houston, TX 77204-3010, USA*
jfparis@uh.edu

Thomas Schwarz, SJ
*Department of Computer Science*
*Marquette University*
*Milwaukee, WI 53233, USA*
thomas.schwarz@marquette.edu

## *Extended Abstract*

## I. Introduction

Reed-Solomon codes [1] have been widely used for protecting data against data losses caused by device failures. For instance, the BackBlaze cloud backup service relies on a 17-out-of-20 RS code to ensure the integrity of its users' data [2]. They split every incoming file into 17 equal-size shards and calculate 3 parity shards so that the file contents can be reconstructed from any 17 of these 20 shards. The main advantage of this approach is its very low space overhead, as only 3 out of the 20 shards contain redundant data.

While Reed-Solomon codes happen to be the most space-efficient technique to protect the contents of $n$ storage units against the failure of $k$ of them, the same is not true when we consider multiple stripes of $n$ units each with their own $k$ parity units. For instance, two-dimensional RAID arrays [3] [4] [5] [6] have been known to protect their data against more triple and quadruple unit failures than multiple RAID level 6 arrays with the same storage capacity and the same space overhead.

At the same time, extant two-dimensional RAID arrays cannot protect data against all quadruple unit failures. While this was not a major issue ten or twenty years ago, we believe the time has come to investigate two-dimensional arrays that can tolerate all quadruple and most quintuple unit failures without experiencing a data loss.

We investigate here the first two-dimensional RAID array that satisfies these criteria. These arrays start with $n$ RAID level 6 parity stripes with two parity nodes per stripe and place their data nodes at the $n(n-1)/2$ intersections of these $n$ stripes. We evaluate how these arrays handle quintuple and sextuple failures and conclude they tolerate many more quintuple and sextuple unit failures than multiple $n$-out-of-$n+k$ arrays with the same storage capacity and the same space overhead. In addition, we show how to eliminate all fatal quintuple failures by adding two superparity units to the array.

## II. Two-Dimensional RAID Arrays

Two-dimensional RAID arrays, or 2D-parity arrays, were first investigated by Schwarz [3] and Hellerstein et al. [4]. In addition, Lee patented a disk array organization with prompt parity updates in one dimension and delayed parity updates in the second dimension [6].

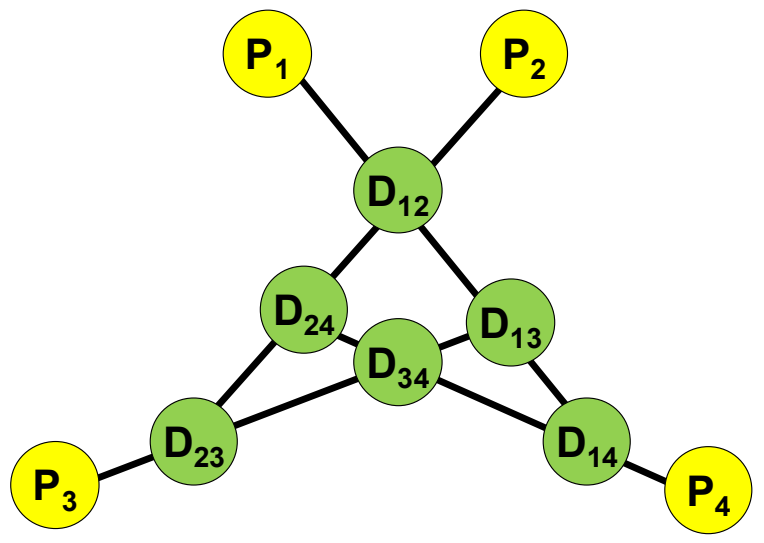

Fig. 1. A small two-dimensional RAID array with six data nodes and four parity nodes.

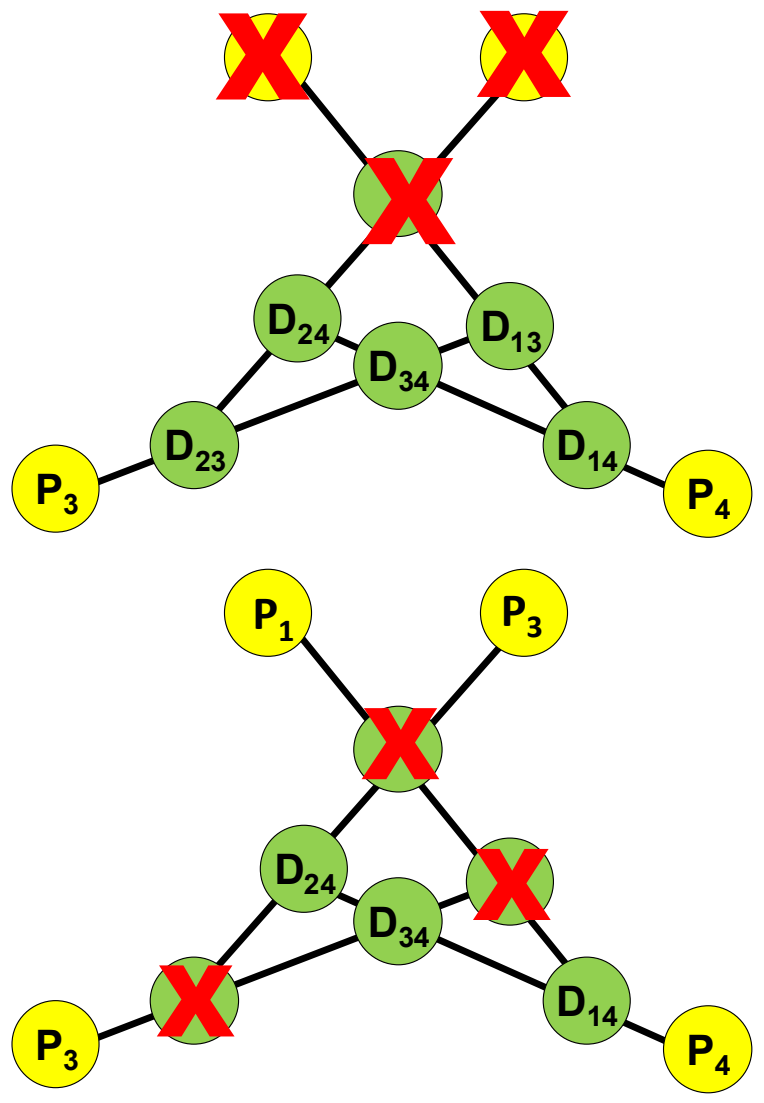

Fig. 2. Two types of triple fatal failures.

Two-dimensional arrays comprise both data nodes and parity nodes. These nodes are organized into parity stripes, each consisting of several data nodes and one parity node containing the exclusive or (XOR) of their contents. In addition,

1. Each data node must belong to two distinct parity stripes.
2. Two distinct parity stripes can only share one data node.

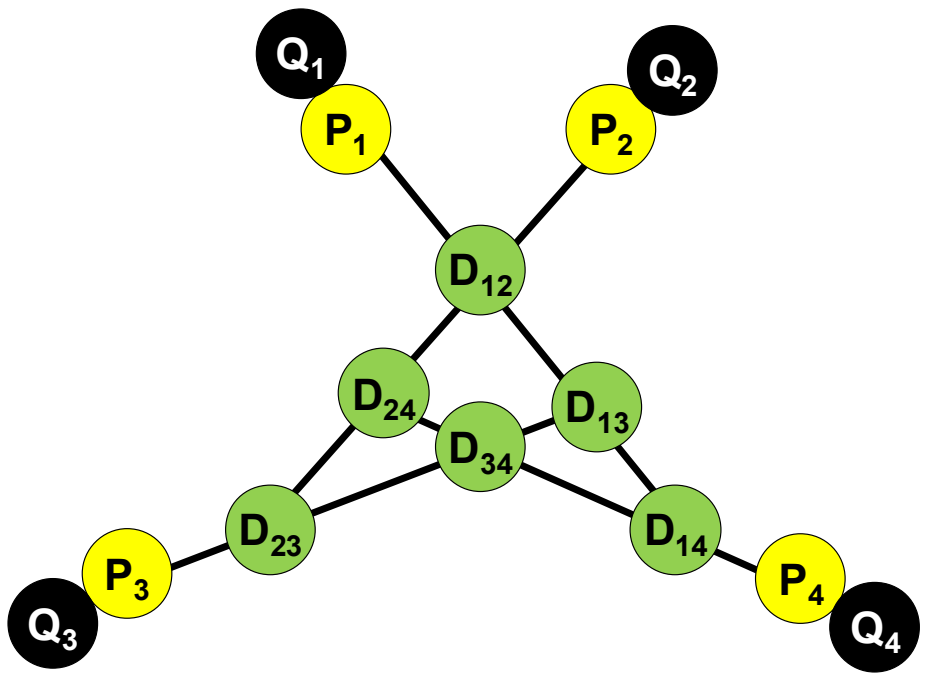

Fig. 3. A small complete two-dimensional RAID level 6 array. Each parity stripe now has two parity nodes.

Consider, for instance, the two-dimensional RAID array displayed in Fig. 1 and note that all its parity stripes intersect each other. As a result, its four parity nodes can protect the contents of six data nodes against all double failures. More generally, any complete two-dimensional array with $n$ parity nodes can contain up to $n(n-1)/2$ data nodes.

While complete two-dimensional RAID arrays are more compact than other two-dimensional RAID array organizations, they are also more susceptible to losing data in the presence of triple node failures. As seen in Fig. 2, complete two-dimensional RAID arrays will experience a data loss whenever:

1. The three failed nodes consist of a data node and the parity nodes of its two parity stripes.
2. The three failed nodes are located at the three intersections of three distinct parity stripes.

This latter vulnerability is the most consequential of the two as it involves more nodes and undergoes a combinational explosion as the size of the array grows. Paris et al. noted the limitation [10] and concluded that the best way to eliminate these failures would be to double the number of parity disks and transform each parity stripe into a RAID 6 level stripe. The idea was not pursued because other two-dimensional RAID arrays could tolerate all triple node failures at a lower cost.

We believe it is time to revisit the topic because we now expect higher levels of fault tolerance than twelve or thirteen years ago. As a result, what was then a mere curiosity now becomes a potential solution for real problems.

## III. Complete Two-Dimensional RAID level 6 Arrays

A complete two-dimensional RAID level 6 array consists of $n$ RAID level 6 parity stripes [3] [8] intersecting with each other and $n(n-1)/2$ data nodes located at the intersection of the $n$ parity stripes.

Each of these $n$ parity stripe contains $n-1$ data nodes as well as two parity nodes for a total of $2n$ parity nodes. To give an example, the small array displayed in Fig. 3 has 4 parity nodes, 6 data nodes, and 3 data nodes per parity stripe. (We expect actual arrays to be much larger and have much lower parity overheads.)

We will now consider the level of data loss protection provided by our organization. Let $N = 2n + n(n-1)/2$ denote the total number of nodes in the array:

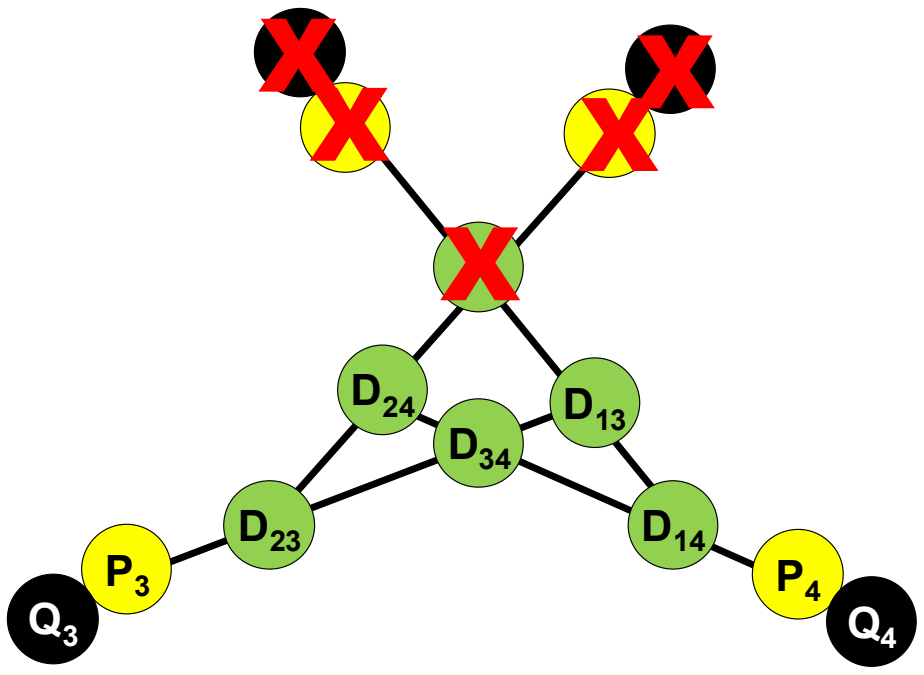

Fig. 4. One of six possible fatal quintuple failures.

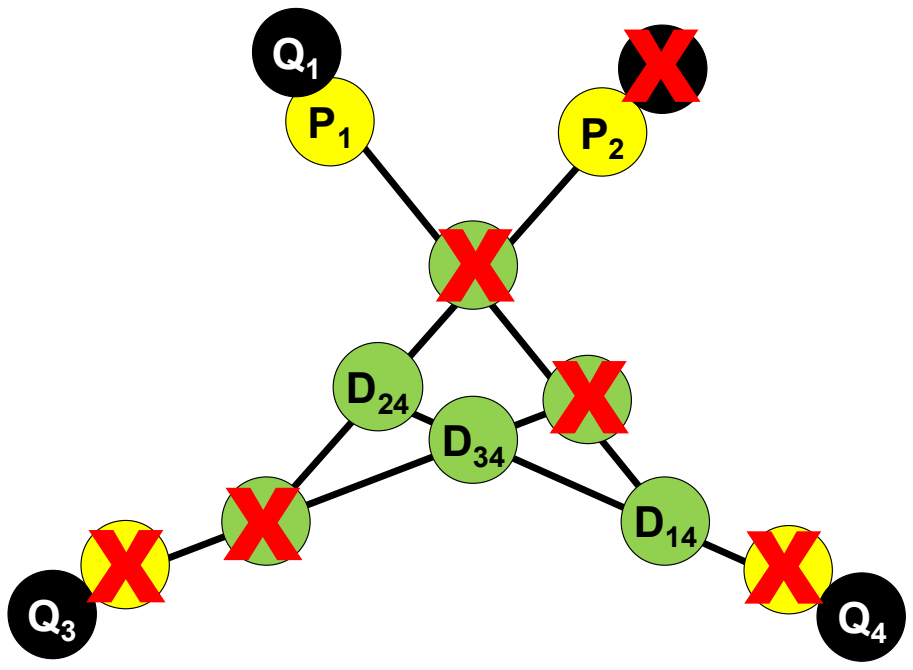

Fig. 5. The fatal failure of three data nodes on the same three parity stripes and one failed parity node in each of these three parity stripes.

1. Complete two-dimensional RAID level 6 arrays tolerate the simultaneous failures of up to four nodes without any data loss.
2. As we can see in Fig. 4, the only quintuple failures that will result in a data loss are the failure of one arbitrary data node and the four parity nodes on its two parity stripes. There are $n(n-1)/2$ such failures out of $\binom{N}{5}$ possible quintuple node failures. As a result, only a fraction
$$\alpha = \frac{n(n-1)}{2\binom{N}{5}}$$
of all quintuple failures will result in a data loss. This fraction quickly decreases as the size of the array increases: less than $2.10^{-6}$ of all quintuple failures are fatal in arrays with more than 45 data nodes and 20 parity nodes.
3. Three different sextuple fault configurations will result in a data loss:
    a. The failure of one arbitrary data node, the four data nodes on its two parity stripes and any other node: there are $(N-5)\, n(n-1)/2$ such fatal sextuple failures.
    b. As shown on Fig. 5, the failure of three data nodes on three distinct parity stripes and one failed parity node in each of these three parity stripes: there are $2^3\binom{n}{3}$ similar fatal sextuple failures.

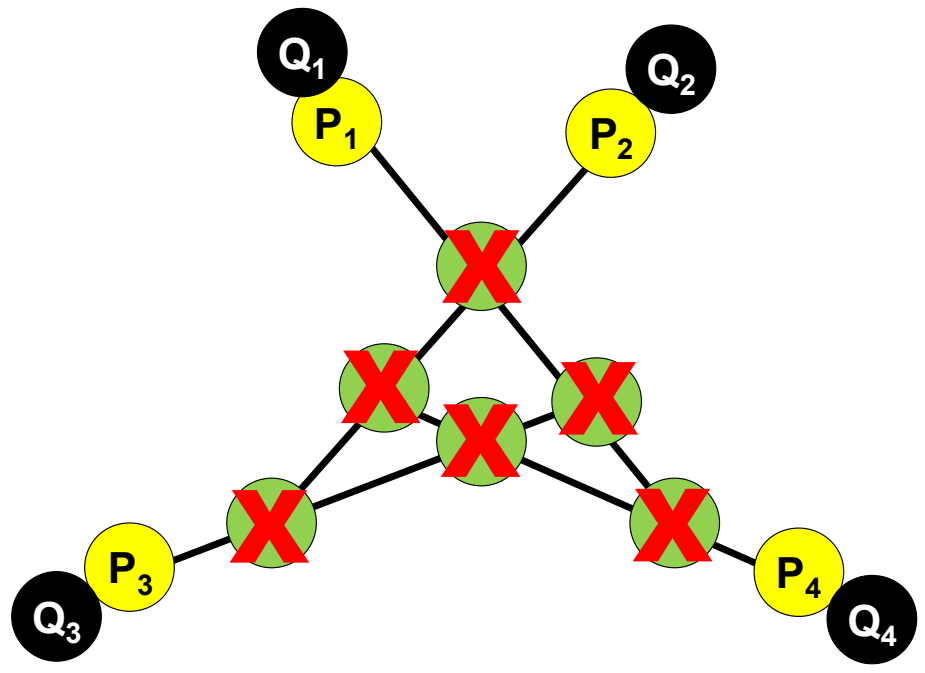

Fig 6. The fatal failure of six data nodes occupying the six intersections of four parity stripes.

c. As shown on Fig. 6, the failure of six data nodes occupying all six intersections of four parity stripes: there are $\binom{n}{4}$ similar fatal sextuple failures.

As a result, the fraction of all sextuple failures that will result in a data loss is:

$$\beta = \frac{(N-5)\frac{n(n-1)}{2} + 2^3\binom{n}{3} + \binom{n}{4}}{\binom{N}{6}}$$

Here too, this fraction quickly decreases as the size of the array increases: less than $3.10^{-5}$ of all sextuple failures are fatal in arrays with more than 45 data nodes and 20 parity nodes.

## IV. A Comparison Study

We wanted to compare the level of data loss protection provided by complete two-dimensional RAID arrays with those achieved by unidimensional $m$-out-of-$m+k$ organizations using the same number of parity nodes to protect the contents of the same number of data nodes against all possible quadruple node failures. We selected an organization consisting of $n/2$ stripes each containing $n-1$ data nodes and four parity nodes. Each stripe thus constitutes a separate $n$–1-out-of-$n$+3

This organization will tolerate without any data loss:

1. The simultaneous failures of up to four nodes.
2. The simultaneous failure of five nodes except when the five failed nodes are all located in the same stripe. There are $\frac{n}{2}\binom{n-1}{5}$ such failures out of $\binom{N}{5}$ possible quintuple node failures. As a result, only a fraction

$$\alpha' = \frac{n\binom{n-1}{5}}{2\binom{N}{5}}$$

of all quintuple failures will result in a data loss.

3. The simultaneous failure of six nodes except when five of the six failed nodes are all located in the same stripe. There are $\frac{n}{2}\binom{n-1}{5}(N-5)$ failures out of $\binom{N}{6}$ possible sextuple node failures. As a result, only a fraction:

$$\beta' = \frac{n\binom{n-1}{5}(N-5)}{2\binom{N}{6}}$$

of all sextuple failures will result in a data loss.

Table I. Compared numbers of fatal quintuple and sextuple failures of the two organizations

| Number of nodes | Fatal quintuple failures | | Fatal sextuple failures | |
|---|---|---|---|---|
| | 2-D RAID 6 | 1-D codes | 2-D RAID 6 | 1-D codes |
| 14 | 6 | 42 | 87 | 378 |
| 27 | 15 | 378 | 465 | 8,316 |
| 44 | 28 | 1,848 | 1,498 | 72,072 |
| 65 | 45 | 6,435 | 3,630 | 386,100 |
| 90 | 66 | 18,018 | 7,425 | 1,531,530 |
| 119 | 91 | 43,316 | 13,559 | 4,938,024 |
| 152 | 120 | 93,024 | 22,820 | 13,674,528 |
| 189 | 153 | 183,141 | 36,108 | 33,69,7944 |
| 230 | 190 | 336,490 | 54,435 | 75,710,250 |
| 275 | 231 | 584,430 | 78,925 | 157,796,100 |
| 324 | 276 | 968,760 | 110,814 | 309,034,440 |
| 377 | 325 | 154,3815 | 151,450 | 574,299,180 |

Table I presents the results of our comparison. As we can see, the unidimensional organization has significantly more fatal quintuple and sextuple failures than our complete two-dimensional RAID level 6 array. In fact, our scheme experiences between 87 and 99.99 percent fewer fatal quintuple failures and between 77 and 99.98 percent fewer fatal sextuple failures than the unidimensional organization does.

## V. Possible Extensions

Since all fatal quintuple failures involve parity nodes, adding superparity nodes provides an easy cost-effective way to protect the array contents against all fatal quintuple failures.

As their name suggests superparity nodes act as parity nodes for regular parity nodes [7]. In our case, we will use them to allow the array to recover from a failure of two pairs of parity nodes. The most cost-effective way to achieve this goal is to connect all P parity nodes or all Q parity nodes (but not both) in a $n$-out-of-$n$+2 stripe. Fig. 7 shows one such organization.

Adding two superparity nodes to the array will also dramatically reduce the number of fatal sextuple failures. Recall that three possible sextuple configurations were responsible for all data loss cases in the presence of six failures:

1. The sole fatal sextuple failures involving one arbitrary data node and the four parity nodes on its two parity stripes must also involve one of the two superparity nodes. There are $n(n-1)$ such sextuple failures.

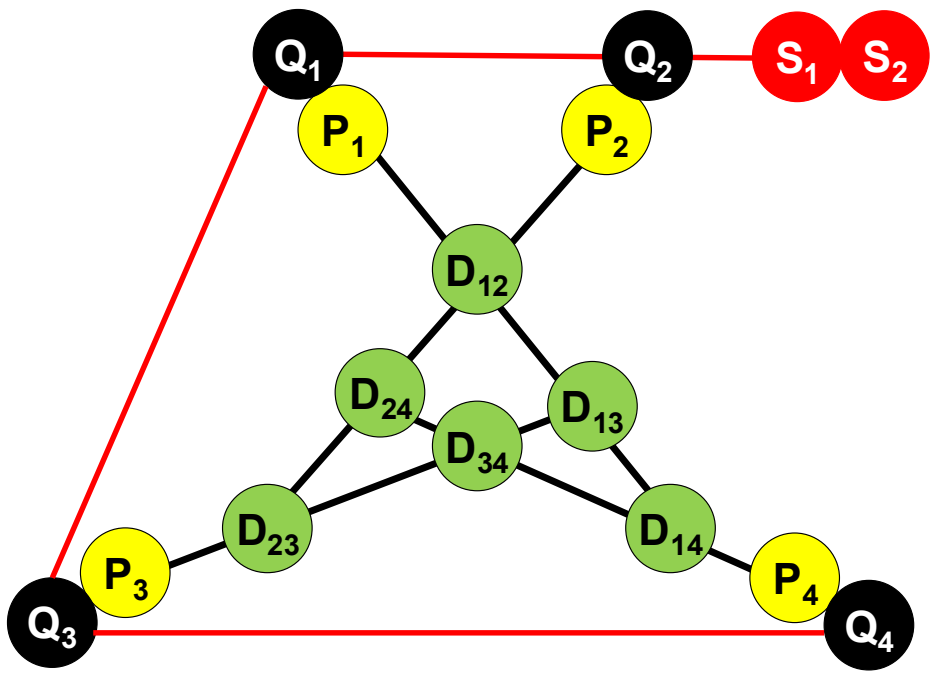

Fig. 7. A small complete two-dimensional array with two superparity devices, $S_1$ and $S_2$, protecting the contents of parity nodes $Q_1$ to $Q_4$.

2. There are only $2\binom{n}{3}$ possible fatal sextuple failures involving three data nodes on the same three parity stripes and one failed parity node in each of these three parity stripes because failed parity nodes must now be either all P parities or all Q parities.
3. The failures of six data nodes occupying six intersections of four parity stripes remain fatal failures.

As a result, the fraction of all sextuple failures that will still result in a data loss is:

$$\beta_{sp} = \frac{2\binom{n}{2} + 2\binom{n}{3} + \binom{n}{4}}{\binom{N_{sp}}{6}}$$

with $N_{sp} = 2n + n(n-1)/2 + 2$.

As Table II shows, adding two superparity nodes to the array eliminates 85 percent of all fatal sextuple failures for all arrays with more than 46 nodes. Adding more superparity nodes would only have a marginal impact on the array resilience as it would still leave the array vulnerable to the simultaneous failures of six data nodes occupying six intersections of four parity stripes.

The most obvious way to eliminate these fatal sextuple failures would be to replace all the array RAID 6 level stripes by $m$-out-of-$m$+3 stripes. The new array would have $n(n-1)/2$ data nodes and $3n$ parity nodes. It would tolerate the failure of up to six of its $3n + n(n-1)/2$ nodes without data loss but would be still subject to some fatal sextuple failures, namely the failures of a data node and the six parity nodes on its two data stripes.

## VI. Conclusion

We have analyzed the performance of the first two-dimensional RAID arrays tolerating quadruple failures. These arrays start with $n$ RAID level 6 parity stripes with two parity sections of these $n$ stripes. Our study has shown that these arrays experience between 87 and 99.99 percent fewer fatal quintuple failures and between 77 and 99.98 fewer fatal sextuple failures than multiple $n$-out-of-$n$+$k$ arrays with the same storage capacity and the same space overhead. In addition, adding two superparity nodes to the array eliminates all fatal quintuple failures and 85 percent of all sextuple failures.

Table II Compared numbers of fatal sextuple failures with and without superparity nodes.

| W/o superparity nodes | | With superparity nodes | |
|---|---|---|---|
| Number of nodes | Fatal sextuple failures | Number of nodes | Fatal sextuple failures |
| 14 | 87 | 16 | 21 |
| 27 | 465 | 29 | 85 |
| 44 | 1,498 | 46 | 238 |
| 65 | 3,630 | 67 | 540 |
| 90 | 7,425 | 92 | 1,067 |
| 119 | 13,559 | 121 | 1,911 |
| 152 | 22,820 | 154 | 3,180 |
| 189 | 36,108 | 192 | 4,998 |
| 230 | 54,435 | 232 | 75,05 |
| 275 | 78,925 | 277 | 10,857 |
| 324 | 110,814 | 326 | 15,226 |
| 377 | 151,450 | 379 | 20,800 |

More work is still needed to evaluate how many fatal septuple failures the array can tolerate either through complete enumeration or through a Monte-Carlo method.